\documentstyle[emulateapj,onecolfloat]{article}

\newcommand{\beq}{\begin{equation}}
\newcommand{\eeq}{\end{equation}}
\newcommand{\tskip}{}

\def\beqa{\begin{eqnarray}}
\def\eeqa{\end{eqnarray}}

\newcommand{\gsim}{\gtrsim}

\begin{document}
\twocolumn[
\title{A Model for Structure Formation Seeded by Gravitationally
Produced Matter}

\author{Wayne Hu$^1$ and P. J. E. Peebles$^2$}
\affil{{}$^1$Institute for Advanced Study, Olden Lane, Princeton, NJ
08540\\ {}$^2$Joseph Henry Laboratories, Princeton University,
Princeton, NJ 08544}
 
\begin{abstract}
This model assumes the baryons, radiation, three families of
massless neutrinos, and cold dark matter were mutually
thermalized before the baryon number was fixed, primeval
curvature fluctuations were subdominant, and homogeneity was
broken by scale-invariant fluctuations in a new dark matter
component that behaves like a relativistic ideal fluid. The fluid
behavior could follow if this new component were a single scalar
field that interacts only with gravity and with itself by a pure 
quartic potential. The initial energy distribution could follow
if this component were gravitationally produced by inflation.
The power spectra of the present distributions of mass and
radiation in this model are not inconsistent with the
measurements but are sufficiently different from the adiabatic
cold dark matter model to allow a sharp test in the near future.
\end{abstract}
]

\section{Introduction}
One motivation for the search for alternative models for
structure formation is that we arrived at the commonly discussed
adiabatic cold dark matter (aCDM) picture after just a
few false starts. This might be because the early universe is
simple enough that there are only a few ways structure could have
originated, or because we were lucky, or perhaps because more
than one model is viable at the present level of constraints. It
seems prudent to continue the search for viable alternatives
before we learn whether they are needed.

The candidate presented here draws elements from a
phenomenological model (Hu 1999) that allows an acceptable
fit to the measured power spectra of
distributions of matter and radiation, and a model with a 
physical provenance within the inflation scenario (Peebles 1999a)
but a poorer fit to the measurements. Our new picture has the 
observational advantage of the former and a physical basis 
that simplifies the latter. It has elements in common with the
phenomenological models analyzed by Bucher, Moodley, \& Turok
(1999), but crucial differences that make the present model
viable.  

We start with the idea that, since the dark matter may interact
only weakly with ordinary matter and radiation, some or all of it may
interact only with gravity (Peebles \&\ Vilenkin 1999a; 
1999b, hereafter PVb). Such
dark matter would be gravitationally produced, as a squeezed
state, by inflation (Ford 1987; Grishchuk \& Sidorov 1990).

We discuss initial conditions from inflation in the next section,
evolution of the departures from homogeneity in \S 3, and tests
of the power spectra of the matter and radiation in \S 4.  

\section{Initial Conditions}

The dynamical components are the cosmic microwave background
(the CMB), three families of massless 
neutrinos, baryons, cold 
dark matter (CDM), and a new dark
component that acts like an ideal fluid with the equation of 
state $p_y=\rho_y /3$.
The primeval energy density contrasts 
satisfy  
\beq
\delta _\gamma = \delta _\nu = {4\over 3} \delta _b = 
{4\over 3}\delta _c = - \delta _y\rho _y/(\rho - \rho _y).
\label{eq:initial_conditions}
\eeq
The last part
expresses the isocurvature condition, where $\rho$
is the total mass density in which the baryons and CDM are initially 
subdominant. The power spectrum of $\delta _y$ is nearly 
scale-invariant: $k^3 P_y(k)$ is initially close to constant.
The first part of equation~(\ref{eq:initial_conditions}) 
says the fluctuations in the usual matter components 
are adiabatic. This can follow if
the chemical potentials of the neutrinos and CDM vanish
and all these components are in mutual thermal equilibrium
that is broken after the baryon number is frozen and before
the CDM is nonrelativistic.
The relativistic 
fluid behavior of $\rho _y$ can follow
from a field that interacts only with gravity 
and with itself by a quartic potential, with the action
\beq
S=\int a^3d^3x\, dt\, (y_{,i}y^{,i}/2 -\lambda y^4/4),
\label{eq:action}
\eeq
and energy density $\rho_y=\dot y^2/2 +(\nabla y)^2/2 +\lambda y^4/4$.

When the frequency of the field oscillation is large compared to
the Hubble parameter $\dot a/a$, equation~(\ref{eq:action})
expressed in conformal time $\tilde t=\int dt/a(t)$ is the
action in Minkowski coordinates for $\tilde y=ay$.
Since the energy of $\tilde y$ in Minkowski spacetime is
conserved, the mean energy density in $y$ scales as 
$\rho _y\propto a(t)^{-4}$ (Ford 1987, Peebles 1999b).  
This means $\rho _y$ can be large 
enough to serve as a primeval seed for structure formation but
remain small enough not to interfere with the standard
models for light element production and gravitational structure
formation. 

Fluctuations in $\rho _y$ are well approximated as linear
acoustic waves from the end of inflation, when
the field starts oscillating, through the characteristic acoustic
oscillation time divided by the density contrast (Peebles 1999b).
The acoustic wave model fails with the appearance of features
that resemble shock waves. If the scale-invariant spectrum of $y$
extends to small wavelengths these shock-like features appear
well before the field fluctuations of interest to astronomy
appear at the Hubble length. The analysis in Peebles (1999b)
indicates that this does not spoil the acoustic wave model
on larger scales.

If the field $y$ in equation~(\ref{eq:action}) exists it will
have been excited, with a near 
scale-invariant spectrum, by inflation (Ford 1987). Kofman \&\
Linde (1987) considered the near  
classical evolution of $y$ in inflation when the potential for
the inflaton $\phi$ also is quartic, 
$V(\phi )=\lambda _\phi\phi ^4/4$. We assume this same
eternal inflation model.

We assume the dimensionless
parameter in equation~(\ref{eq:action}) is in the range 
\beq
\lambda _\phi\ll\lambda < 0.01 .
\label{eq:lambdas}
\eeq
The lower bound makes the $y$-field energy subdominant to the
inflaton during inflation (PVb); otherwise
$y$ assumes the role of the inflaton (Felder, Kofman \&\  
Linde 1999). The upper bound is chosen so $y$ is close to
constant across the present Hubble length. This follows from
a consideration of the competition between the freezing of
quantum fluctuations that tend to drive the field value away from
zero and classical dissipation as the field rolls to
the minimum of its potential at $y=0$. Early in inflation these
processes are close to statistical equilibrium.
As the value of the 
Hubble parameter $H$ decreases
equilibrium eventually is broken, at expansion parameter
$a_e$, and $y$ thereafter evolves almost as a classical field.
Under the upper bound in equation~(\ref{eq:lambdas}) 
the expansion from $a_e$ to the end of inflation is large enough
that $y$ is close to constant across our Hubble
length.\footnote{This follows by adding numerical factors
to equation~(33) in PVb, to get the
$y$-field relaxation time 
$\tau _y =\pi\sqrt{6}H^{-1}\lambda ^{-1/2}$, with
$\tau _y = H_x^{-1}$ at $a=a_e$. This with the expansion factor
$a_x/a_p\sim e^{70}$ 
to the end of inflation from the time of freezing of the
fluctuations we see, with the condition $a_e<a_p$, fixes
the bound on $\lambda$.} We assume that at our
position the field value at $a=a_e$ is close to the
characteristic value at equilibrium,
$\lambda \left< y^4 \right> = 3H^4/8\pi ^2$ (Starobinsky \&
Yokoyama 1994).  
The perturbations to $y$ added from $a=a_e$ to the end of
inflation produce a near Gaussian scale-invariant fluctuation
spectrum with variance per logarithmic interval of wavenumber
(PVb eq.~[39])
\beq
\left(\delta y\over y\right) ^2 \approx 
{\lambda\over 6\pi ^2\log _ea_x/a_p}.
\label{eq:delta_y}
\eeq
As discussed in the next section the large-scale perturbation to
the present thermal background radiation (the CMB) is 
$\delta T/T = (\delta y/y)/5f$, 
where $f$ is the ratio of mass densities in radiation and
neutrinos to the energy in the $y$-field. 
The fit to the observed temperature variance per logarithmic
interval of $l$, $(\delta T/T)^2=(1\times 10^{-5})^2$, requires
\beq
\lambda \approx 1\times 10^{-5}f^2.
\label{eq:limit}
\eeq
The value of $f$ depends on the model for the origin of ordinary
matter and radiation; we must treat $f$ as an adjustable
parameter. The standard model for the light elements requires
$f\gsim 10$, leaving a small window of consistency between
equations~(\ref{eq:lambdas}) and~(\ref{eq:limit}). 

\section{Evolution}

The evolution of the distributions of matter and radiation in
linear perturbation theory is computed by the usual methods.   
We discuss only aspects that differ from the usual
case.\footnote{Though there is no gauge ambiguity in 
our initial conditions numerical stability in the evolution 
requires careful choice of gauge and metric variables (Hu 1999).}

The dynamics of the fluctuations are governed by two events: the
transition from radiation- to matter-dominated expansion, and
Hubble crossing, when relativistic stress gradients 
and gravity have comparable dynamical effects. We consider first a
large-scale mode that crosses the Hubble length after matter-radiation
equality.   
The residual entropy fluctuation,
\beq
\sigma = \delta_c - {3 \over 4}\delta_{\rm rel} \approx \delta_c = 
	{3 \over 4}\delta_\gamma \,
\eeq
where $\rho_{\rm rel} =\rho_\gamma+\rho_\nu +\rho_y$, 
becomes important near matter-radiation equality and before
pressure causes the mode to oscillate. On these large scales the
relative distribution of the $y$-field and the
familiar radiation species is irrelevant because their
gravitational effects exactly cancel before Hubble crossing and
are  negligible afterward. The relation between the present
distributions of the cold dark matter and the CMB shows an
interesting effect. The gravitational potential in the
matter-dominated regime is related to the initial entropy
fluctuation as $\Phi = {\sigma /5}$, and $\Phi$ is related to the
density perturbation by the Poisson equation (Kodama \& Sasaki
1986). The CMB anisotropy due to the gravitational
redshift is 
\beq
{\delta T \over T}\Big|_{\rm grav} = 
		-2\Delta \Phi + {\delta T \over T}\Big|_{\rm init}
	    = -{1 \over 3}\Phi \,.
\label{eq:SW}
\eeq
This is the same as the adiabatic CDM (aCDM) model, and
different from the isocurvature CDM (iCDM) model  
in Peebles (1999a) where primeval fluctuations in the CDM are
balanced by the CMB and neutrinos and 
$\delta T/T |_{\rm grav} = -2\Phi$. It is
helpful to the construction of a viable model 
to assume 
the initial photon distribution follows that of the species
that is responsible for gravitational structure formation in the
matter-dominated epoch (Hu 1999). This puts the photons that
initially are hottest where the gravitational potential becomes the
deepest, so the temperature fluctuation is reduced 
as the photons move out of the potential 
well. 
The consequence is that in aCDM and the present model the 
observed ratio of matter to radiation fluctuations follows 
from a near scale-invariant primeval fluctuation spectrum, 
while in the iCDM model a fit to this ratio requires a substantial
tilt to increase small-scale fluctuations over large.  

\begin{figure}[t]
\centerline{\epsfxsize=3.5truein\epsffile{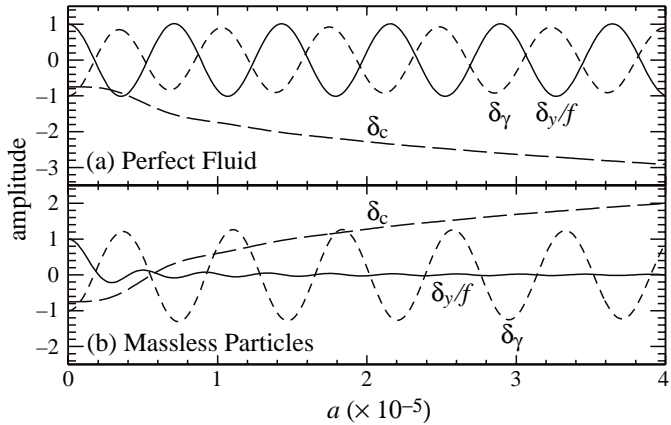}}
\caption{Time evolution of Fourier amplitudes in a mode with very
short comoving wavelength for (a) our perfect fluid model for the
$y$-component and (b) a free massless particle model. Note the
zero crossing of the CDM 
amplitude $\delta _c$ in (b). In a mode at
$k\sim 0.15h$~Mpc$^{-1}$, 
$\delta _c$ crosses zero near matter-radiation equality,
producing the zero of $P(k)$ in the dashed line in Fig.~2.}
\label{fig:time}
\end{figure}

The evolution of small-scale fluctuations in the CDM 
that cross the Hubble length before matter-radiation equality is
surprisingly sensitive to the behavior of the new $y$-component.
At Hubble crossing, stress gradients in the relativistic
components cause them to oscillate. The CMB density oscillates  
as an acoustic (sound) wave with amplitude given by the
initial conditions
in such a way that observationally acceptable peaks 
result from scale-invariant initial conditions. The
$y$-component in our model also   
oscillates as an acoustic wave. Aside from the neutrinos,
this keeps the radiation distributions almost balanced. 
The CDM amplitude appears at the Hubble length after moderate 
growth from the initial value (Fig.~\ref{fig:time}a). 
As in aCDM, the modest further increase 
of $\delta _c$ to matter-radiation equality leaves the usual 
$k^{-2}$ suppression of
small-scale power and an observationally acceptable present
mass fluctuation spectrum from initially scale-invariant
fluctuations. If instead the $y$-component were a free gas of
massless particles, free streaming would  
cause an imbalance with the acoustic oscillation of the
CMB, as in Figure~\ref{fig:time}b.  The 
resulting metric perturbation reverses the sign of $\delta _c$.
The reversal can only be produced during the
radiation-dominated epoch, so there is a mode near the
Hubble scale at matter-radiation equality that is caught in the
act of reversal, producing a zero in the linear power spectrum, 
and spoiling the fit to large-scale structure measurements. 

\begin{figure}[t]
\centerline{\epsfxsize=3.5truein\epsffile{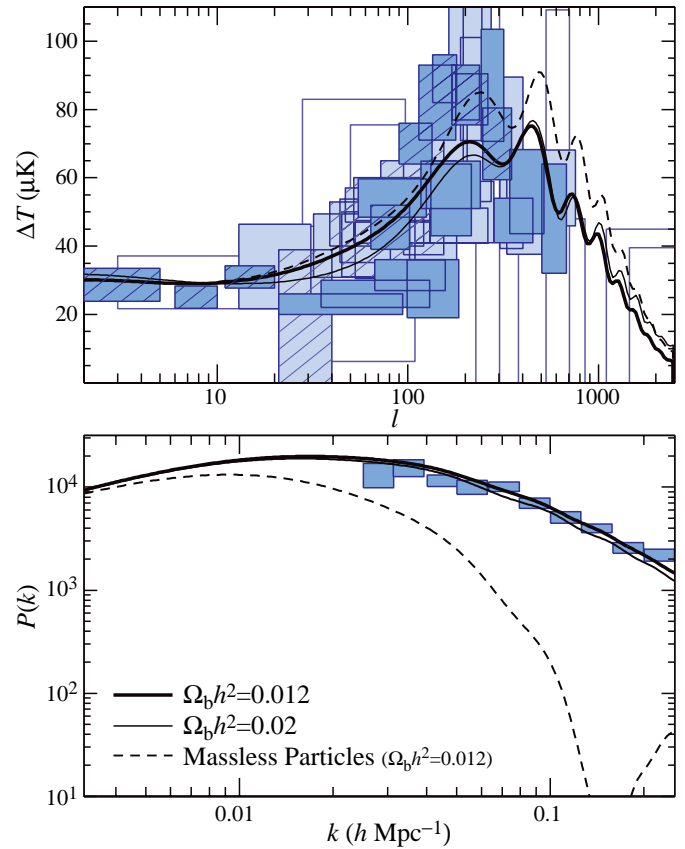}}
\caption{Measured and modeled power spectra of the CMB and
galaxy distributions. The shading of the CMB
error boxes is according to area and corresponds
to the $1\sigma$ errors $\times$ the width of the experimental 
window; cross hatching denotes measurements included in set ``A''. 
Our models are plotted as solid
curves; the dashed curve represents a model where the seed field
is a gas of free 
massless particles 
rather than a fluid.} 
\label{fig:power}
\end{figure}

\section{Phenomenology}

Figure~\ref{fig:power} compares our model predictions to all
significant measurements of the CMB temperature anisotropy and to
the Peacock \& Dodds (1994) compilation of measurements of the
power spectrum of the galaxy distribution. We assume a
scale-invariant initial spectrum of fluctuations in $y$ 
(as in eq.~[\ref{eq:delta_y}]), 
standard recombination,
$\Omega_m=0.35$, $\Omega_\Lambda=0.65$, 
$h=0.8$,  and two values of the
baryon density, $\Omega_b h^2=0.02$ and $\Omega_b h^2=0.012$. The
former is close to the central value of Burles et al. (1999)
based on the deuterium abundance 
${\rm D/H} = 3.4\times 10^{-5}$.
Kirkman et al. (1999) consider the most secure bound to be
${\rm D/H} < 6.7\times 10^{-5}$; this abundance scales the baryon
density to the lower number.  

Following Tegmark (1999) and Miller et al.\ (1999),
we 
find a crude estimate of $\chi^2$
for the CMB temperature anisotropy by treating all data
points as independent with Gaussian distributions of errors.
The first two lines of 
Table 1 list the reduced $\chi^2$ employing all the
data (58 points), the selection 
in Miller et al.~(1999)
(``A''; 24 points),  and the remaining data plus the COBE DMR results 
(``B''; 42 points).
The third line is the best fit aCDM model from 
Tegmark (1999). 
Values for the full data set and selection ``B'' are arguably less secure because 
they are based on a more heterogeneous set of methods. 
The $\sim 10\%$ calibration uncertainty, which is not included in these 
$\chi ^2$ estimates, is a serious general barrier
to the interpretation in terms of formal
measures of significance. Within 
the calibration uncertainty our
low baryon density model seems viable, although challenged by the
D/H measurements (Kirkman et al.\ 1999), while our high density model
is 
challenged but 
we believe not ruled out by the CMB measurements. 

\begin{center}
{TABLE 1\\[4pt] \scshape Approximate CMB $\chi^2/\nu$} \\[3pt]
\begin{tabular}{llll}
\tskip\tableline\tableline\tskip
Model & All & A & B\\
\tableline\tskip
$\Omega_bh^2=0.012$ & 2.6 & 1.5 & 1.3\\
$\Omega_bh^2=0.02$  & 2.7 & 2.0 & 1.3\\
aCDM 		    & 2.5 & 1.2 & 1.4\\
\tskip\tableline\tskip
\end{tabular}
\end{center}

We get satisfactory agreement with 
the power spectrum of the galaxy distribution. The 
normalization implies $\sigma_8=0.84-0.86$, consistent with 
the bounds $0.74 \la \sigma_8 \la 1.1$ implied by the abundance
of rich clusters of galaxies at our model parameters (e.g. Viana
\& Liddle 1999). 

Our model differs from those of
Bucher et al.\ (1999) in two ways. First, the CDM
density perturbations follow the CMB 
(eq.~[\ref{eq:initial_conditions}]). 
We noted that this suppresses 
the large-scale anisotropy of the CMB, allowing a near
scale-invariant primeval spectrum and making the 
peaks in the anisotropy spectrum at $\ell >100$ more prominent.
The same effect follows from a 
{\it coherent} superposition of the CDM-isocurvature and neutrino
isocurvature modes of Bucher et al.\ (1999); one 
cannot use a linear combination of the individual
power spectra.  Second, motivated by equation~(\ref{eq:action}),
we model the isocurvature departure from 
homogeneity by a component that behaves as a perfect
fluid rather than a gas of free massless particles. One sees
in Figure~1 and the dashed curves in Figure~2 that the 
CMB anisotropy is not much affected but the free particle model
produces a zero in the mass power spectrum (in linear
perturbation theory) at an undesirable wavelength.  

The power spectra of the present distributions of matter and
the CMB depend on the cosmological parameters in different
ways from the aCDM model. The heights of the peaks
in the CMB spectrum depend on $\Omega_b h^2$ in opposite
ways (Hu \& White 1996): here the 
odd-numbered peaks represent rarefaction of the photon fluid in
the potential wells and hence decrease when the baryon density is
increased, as one sees in Figure~2. Also, since our model has 
no initial metric fluctuations whose decay in the
radiation-dominated epoch enhance the peaks, 
the peak values do not increase with decreasing $\Omega_m h^2$.  
The lesson here is a general one:  cosmological parameters
derived from a model fit are provisional no matter how securely
fixed within the model until the model is unambiguously
established.  

\section{Discussion}

We conclude that our model is viable 
but likely to be critically tested by 
CMB anisotropy measurements in progress. The same is true of the
aCDM model, of course.  

Our model can be adjusted; here are four considerations.
First, we use isocurvature initial conditions. There may be a
significant adiabatic perturbation from the inflaton, or, in other
inflation models, from the stress of the $y$-field fluctuations
during inflation. Second, we place $\lambda$ in a narrow window 
(eqs.~[\ref{eq:lambdas}] and~[\ref{eq:limit}]). 
If $\lambda\gsim 0.01$ the fluctuations in $\rho _y$ at the end of
inflation have positive skewness, so the primeval fluctuations
in the CDM mass distribution are non-Gaussian with negative
skewness. Models with positive skewness are seriously constrained
(Frieman \&\ Gazta\~naga 1999); negative
skewness may be interesting for structure formation. The second
moments needed for the tests in Section~4 have not been analyzed
for this case. Third, one may ask whether some or all of the CDM
is in fields that interact only with gravity and themselves by
potentials that would have to be more complicated than the
quartic considered here. PVb present preliminary elements of a
model for this more complicated case. 
Fourth, we have assumed standard recombination. One could imagine
stars present at $z\sim 1000$ delay the rapid drop in ionization;
that would shift the peaks in the CMB spectra to smaller $\ell$
and substantially change the significance of this test.

The field $y$  is a new hypothesis. 
Its parameter $\lambda$ is not exceedingly small, however, and by
moving the seed for structure formation from the inflaton we
remove the requirement for a specific value of the very small
parameter $\lambda _\phi$. But closer consideration
of these issues might best await observational developments.  

It may be significant that the structure formation
history in our model is only mildly different from aCDM. 
Perhaps this is telling us viable phenomenological models
already are limited: they have to approximate
aCDM. Or perhaps our imagination in exploring concepts like
gravitationally produced matter is limited.   

\acknowledgements

We are grateful to Max Tegmark for the use of his code to
calculate our estimates of reduced $\chi ^2$, Lyman Page for
guidance to the CMB measurements, and David 
Tytler for guidance to the D/H constraint. We also benefited from
discussions with Andrei Gruzinov, Lam Hui, Alexander Vilenkin,
 and Matias Zaldarriaga.  
W.H. was supported by the Keck Foundation, a Sloan Fellowship and
NSF-9513835.  P.J.E.P. was supported in part at the Institute 
for Advanced Study by the Alfred P. Sloan Foundation and at
Princeton University by the NSF.


\begin{thebibliography}{99}

\bibitem[Bucher et al.]{BMT}
Bucher, M., Moodley, K. \&\ Turok, N. 1999, astro-ph/9904231

\bibitem[Burles et al.]{Burles}
Burles, S.,  Nollett, K.M., Truran, J.N., \& Turner, M.S., 1999
PRL, 82, 4176 

\bibitem[Felder et al.]{fkl}
Felder, G., Kofman, L. \&\ Linde, A. D. 1999
hep-ph/9903350

\bibitem[Ford]{Ford}
Ford, L. H. 1987, Phys Rev, D35, 2955 

\bibitem[Frieman]{Fr} Frieman, J. \&\ Gazta\~naga. E. 1999, 
astro-ph/9903423

\bibitem[Grishchuk \& Sidorov]{Grisch}
Grishchuk, L.P. \&\ Sidorov, Y. V. 1990, 
Phys Rev D, 42, 3413

\bibitem[Hu]{Hu}
Hu, W. 1999, Phys Rev, D59, 021301

\bibitem[Hu \& White]{HuWhi}
Hu, W. \& White, M. 1996, ApJ, 471 30

\bibitem[Kirkman et al.]{Kirk} Kirkman, D., Tytler, D., Burles,
S., Lubin, D., and O'Meara, J. M. 1999, astro-ph/9907128

\bibitem[Kodama \& Sasaki]{KS}
Kodama, H. \& Sasaki, M. 1986, Int. J. Mod. Phys., A1, 265

\bibitem[Kofman \& Linde]{KL} 
Kofman, L. \&\ Linde, A. D. 1987, 
Nucl Phys, B282, 555

\bibitem[Miller]{Miller} Miller, A. D. et al. 1999, 
ApJ, 524, L1

\bibitem[Peacock \& Dodds]{PD}
Peacock, J.A. \&\ Dodds, S.J. 1994, MNRAS, 267
	1020

\bibitem[Peebles]{Peeb99a}
Peebles, P. J. E. 1999a, ApJ, 510, 531

\bibitem[Peebles]{Peeb99b}
Peebles, P. J. E. 1999b, preprint

\bibitem[Peebles \& Vilenkin]{PVa}
Peebles, P. J. E. \&\ Vilenkin, A. 1999a, Phys
Rev, D59, 063505

\bibitem[Peebles \& Vilenkin]{PVeb}
Peebles, P. J. E. \&\ Vilenkin, A. 1999b, Phys
Rev, in press; PVb

\bibitem[Starobinsky \& Yokoyama]{SY}
Starobinsky, A. A. \&\ Yokoyama, J. 1994, Phys Rev D50, 6357

\bibitem[Tegmark]{Tegmark}
Tegmark, M. 1999, ApJL, 514, 69

\bibitem[Viana \& Liddle]{VL}
Viana, P.T.P \&\ Liddle, A. 1999, MNRAS, 303, 535

\end{thebibliography}
\end{document}